\documentclass[aps,prb,twocolumn,superscriptaddress,floatfix]{revtex4}
\usepackage{graphicx}
\usepackage{bm}

\begin{document}

\title{Exact solution for many-body Hamiltonian of interacting particles with linear spectrum }

\author{M.V. Entin}
\affiliation{Rzhanov Institute of Semiconductor Physics, Siberian
Branch of the Russian Academy of Sciences, Novosibirsk 630090, Russia}
\affiliation{Novosibirsk State University, Novosibirsk 630090, Russia}
\email{entin@isp.nsc.ru}

\author{L. Braginsky}
\affiliation{Rzhanov Institute of Semiconductor Physics, Siberian
Branch of the Russian Academy of Sciences, Novosibirsk 630090, Russia}
\affiliation{Novosibirsk State University, Novosibirsk 630090, Russia}
\email{brag@isp.nsc.ru}

\begin{abstract}
The exact solution of the Schr\"odinger equation for the one-dimensional  system of interacting particles with the linear dispersion law in an arbitrary external field is found. The solution is reduced to  two groups of particles moving with constant velocities in the opposite directions with a fixed distance between the particles in each group. The problem is applied to the edge states of the 2D topological insulator.
\end{abstract}

\maketitle

\subsection*{Introduction}
One-dimensional electron systems with the linear dispersion are the topical  issue  now:  the edge states \cite{qi}-\cite{konig} of the 2D topological insulator \cite{bern}, graphene strips,  carbon nanotubes \cite{miserev1}-\cite{miserev3}, etc. The specificity of these systems is the presence of the double degenerate state at the zero longitudinal momentum that has a linear splitting at a finite momentum.

The 1D systems with near-linear spectrum are the subject of study in the theory of bosonization and Luttinger liquid \cite{tsvelik}. This approach considers the electron Fermi liquid of strongly-interacting electrons  assuming    their energy spectrum  to be approximately linear near the Fermi energy. The bosonization procedure separates the long- ($q\ll p_F$) and short-range ($q\approx 2p_F$) interactions considering them in different ways. In general, this approach is approximative and applicable near the Fermi points only.

Note that the curvature of the energy spectrum caused by the k-p expansion violates the linearity. For example, such non-linear corrections to the energy spectrum exist in graphene where they determine the unconventional character of the e-e scattering \cite{golub} and e-h coupling to the excitons \cite{mahmood}.
The linear spectrum of the 2D TI edge states is the main reason of disappearance of the electron correlation energy \cite{ent-brag}.

In \cite{hyper} we have studied the linearity of the energy spectrum in the edge states of the 2D topological insulator. We concluded that in two models of the edge states \cite{volkov} and \cite{qi}-\cite{konig} the linearity is absolute, while in other cases the non-linear corrections are extremely weak. This pushes forward the problem of the many-electron states in the system with the linear single-electron energy spectrum.

The purpose of the present paper is the general consideration of the one-dimensional system with  the many-body Hamiltonian
\begin{equation}\label{100}
  H=\sum_i (\frac{1}{\hbar}v\sigma_ip_i+U(x_i))+ \sum_{i<j}V(x_i-x_j).
\end{equation}
Here $U(x)$ is an external field, and $V(x_i-x_j)$ is the interaction between the particles, $[x_i,x_j]=0,~[p_i,p_j]=0,~[p_i,x_j]=-i\hbar$. For certainty we consider $\sigma_i=\pm$ (or equivalently, $\sigma_i=\uparrow,\downarrow$) as a spin quantum number.  Below we shall set $\hbar=1$.  The coordinates of spin-up and spin-down electrons are denoted as $x_i$ and $y_i$, accordingly. The Hamiltonian Eq.(\ref{100}) is valid for the edge states of electrons in 2D topological insulators \cite{hyper}.

 The linearity of the energy spectrum is the most important for our consideration. We  obtain an exact solution of this quantum problem.
Unlike the case of the Luttinger liquid, we do not need low temperatures and the vicinity of the energy to the Fermi level. Owning to the exact linearity of the spectrum our results are valid for all energies.

\subsection*{Exact solution of many-body Schr\"odinger equation}
In the absence of interaction the direction of motion coincides with the   spin sign, so that the spin-up and spin-down electrons are rightmovers and leftmovers, correspondingly.
The exact solution of the Schr\"odinger equation for a single particle with energy $E=pv$ is $\psi(x)=\exp(ipx-i\int dx U(x)/v)$. This is the wave function of constant density $|\psi|^2$. The result essentially differs from that for a particle with a quadratic kinetic energy.

Consider now two free ($U(x)=0$) particles with positive spins. The Hamiltonian
\[
H=vp_{1,\uparrow}+vp_{2,\uparrow}+V(x_1-x_2)
\]
 commutes with $x_1-x_2$ conserving, therefore,  the distance between the particles. Then the wave function can be chosen as the eigenfunction of $x_1-x_2$ and the total momentum $P=p_{1,\uparrow}+p_{2,\uparrow}$: $\psi=A\exp(iP(x_1+x_2)/2)\delta(x_1-x_2-a)$, where $a$ is the eigenvalue of $x_1-x_2$. The corresponding selfenergy is
 \[
  E=Pv+V(a).
 \]
 The normalizing coefficient $A$ should be chosen to exclude the divergence of the integral $\int dx \delta^2(x)$. This divergence  inevitably appears when one uses the selffunctions of the coordinate operator. It can be formally fixed by the choice  $A^2=1/\delta(0)$.

For the case of  $n$ rightmovers we write
\begin{equation}\label{200}
  \psi(x_1,....x_n)=A^n\exp(i PX)(\Pi_k\delta(x_{k+1}-x_k-a_k)),
\end{equation}
where $X=\sum_ix_i/n$, $P=\sum_ip_{i,\uparrow}$ is the total momentum. The selfenergy is
\begin{equation}\label{250}
E=Pv+\sum_{i<j} V\left(a_{ij}\right), ~~~a_{ij}=\sum_{i}^{j-1} a_k.
\end{equation}
To include an external, e.g. impurity, potential $U(x_i)$ into consideration, we multiply the wave function Eq.~(\ref{200}) by the factor
\[
u=\exp(-i \int dx_1 \sum_jU(x_1+a_j)).
\]
The antisymmetry of the coordinate part of the wave function can be  achieved by means of the Slatter determinant  composed of the functions $f_{i,j}=\delta(x_i-x_j-a_{i,j})$. Thus, the wave function of the $n$ right- (left-)movers is
\begin{equation}\label{300}
  \psi(x_1,....x_n)=u\exp(i PX)\mbox{det}(f_{i,j}).
\end{equation}

Consider now two particles of the opposite spins with the Hamiltonian:
\[
H=vp_{\uparrow}-vp_{\downarrow}+V(x-y).
\]
 The wave function can be chosen as the eigenfunction of $x+y$ with selfvalue $2c$:
 \begin{eqnarray} \psi=\exp\left(i P'(x-y)/2-i\int dxV(2b-2x)/v\right)\nonumber\times \\A\delta(x+y-2c).
 \end{eqnarray}
 Here $P'=p_1-p_2$, the corresponding selfenergy is $E=P'v$.

 Finally, consider a general case of $n$ spins up and $m$ spins down. The Hamiltonian   commutes with the central point between each leftmover and rightmover $(x_1+y_1)/2$. The sufficient condition is a fixation of one of these variables, e.g., $x_1+y_1=2c$. Hence, the state with the quantum numbers $P,~c,~\{a_k\},~\{b_k\}$ is
\begin{eqnarray}\label{rightleft u}\nonumber
&&|P,c,\{a_1,...a_{n-1}\},\{b_1,...b_{m-1}\}\rangle=\\\nonumber
&&\psi(x_1,..x_n;y_1,...y_m)=\Phi\exp(i P\zeta)A\delta(y_1+x_1-2c)\times\\&&\prod_k^{n-1}A\delta(x_{k+1}-x_k-a_k)\times \prod_k^{m-1}A\delta(y_{k+1}-y_k-b_k)),
\end{eqnarray}
where $\zeta=(x_1-y_1)/2$, $m>1,~n> 1$. If $m=1$ or $n=1$, the corresponding product in Eq.(\ref{rightleft u})should be replaced  by unity.  Substituting into the Schr\"odinger equation, we find the  proportionality factor
\begin{eqnarray}\label{rightleft uu}
&&\Phi=\exp\Big(-\frac{i}{v}\int^{x_1}dx_1({\cal V}(x_1)+{\cal U}(x_1)\Big)\\\nonumber
&&{\cal V}(x_1)=\sum_{i,j}^{n,m}V\Big(2x_1-2c+a_{i,1}-b_{i,1}\Big)\\
&&{\cal U}(x_1)=\sum_{i=1}^nU\Big(x_1+a_{i,1}\Big)-\sum_{i=1}^mU\Big(2c-x_1+b_{i,1}\Big)\nonumber\\
&&a_{i,j}=\sum_{k=i}^{j-1}a_k,~~~b_{i,j}=\sum_{k=i}^{j-1}b_k.\end{eqnarray}
The corresponding energy is
\begin{eqnarray}\label{tot}E=Pv+\tilde{U},~~ \tilde{U}=\sum_{i> j}U\left(a_{i,j}\right)+\sum_{i> j}U\left(b_{i,j}\right).\end{eqnarray}

The case of an arbitrary number of identical electrons with different spins should be considered using the permutation symmetry and
Young diagrams. The ground state of the system corresponds to the equal numbers of up and down spins, so that the  total spin is zero. Note that in this state the average velocity vanishes. Thus, the spin wave function is symmetric with respect to all particles, and the coordinate wave function should be antisymmetric with respect to all coordinates.  To construct  the appropriate coordinate wave function, we need to
antisymmetrize the wavefunction.

Neglecting the particle exchange, each of subsystems of right- and leftmovers represents a
solid superparticle, inside which the distances between
the electrons are fixed. The right  and left moving
superparticles (RMS) and (LMF) obey the linear dispersion. The
Hamiltonian of the  RMS and LMS interaction depends on the
distance between the electrons belonging to the different superparticles. Thus, RMS and LMS
 can be considered as two opposite moving superparticles  and the two-particle wave function can be used to
describe their relative motion. This is an explanation of Eqs. (\ref{rightleft u}) and (\ref{rightleft uu}).

\subsection*{Cyclic boundary conditions}
In the previous consideration we assumed that the coordinates change in the infinite domain $-\infty<x_i<\infty$. In this case the energy is expressed via the total  and relative momenta of the left- and right-movers (\ref{tot}) and the   interaction energies inside the groups. It is important to point out that the interactions between the carriers from different groups as well as the carriers with the impurities do not contribute to the total energy.

Actually, this is consequence of the problem formulation. We have considered an infinite system with the finite number of interacting  electrons.  In this case electrons with different spins being separated at infinite distance  can be characterized  by their momenta at the infinity. In a dense system, however, this is not the case.
Consider now a cyclic system of the length $L$ assuming $x_i+L=x_i$. Suppose the potentials $V(x)$ and $U(x)$ to be the periodic functions of  $L$. For this reason we replace $x$  by the distance between points  on the circle $x\rightarrow(L/2\pi)\sin (2\pi x/L)$. In particular, the $\delta$-functions in the previous expressions have to be replaced by their periodic generalizations $\delta(x)\to (2\pi/L)\delta(\sin (2\pi x/L))$.

The cyclic boundary condition reads $\psi(x_1,...x_i+L,...)=\psi(x_1,...x_i,...)$.  Consider first the
two-particle problem with the opposite spins. In this case the quantization rule Eq.~(\ref{250}) is

\[
E=\frac{2\pi vN}{L}+\frac{1}{L}\int_0^L(V(x)+2U(x))\,dx,
\]
where $N$ is an integer. The second term here is the average interaction, which has to be added to the total energy.
The generalization of the quantization rule to many particles is
\begin{eqnarray}\label{E}
&&E=2\pi vN(n-m)/L+\sum_{\sigma=\pm,i,j}^{n,m}V\left(s_{\sigma,i,j}\right)+\nonumber\\
&&nm\int_0^LV(x)dx+(n+m)\int_0^LU(x)dx.
\end{eqnarray}
 In accordance with Eq.(\ref{E}),  the total energy incorporates  the  intra-group interaction  and averages of the external field and inter-group  interaction. The physical meaning  can be simply understood: electrons with same spins conserve the distance between each other (and, consequently the sum of the potentials in Eq.(\ref{E})) and move through the impurity lattice and the electrons with opposite spins, averaging the interaction with them (the second line in Eq.(\ref{E})).

It is clear from Eq.(\ref{E}) that the electron density is not affected by an external potential. This explains the absence of the correlation energy \cite{ent-brag} for the system with the linear spectrum.

Note that  $P$,~  $a_k,~b_k$ and  $c$ compose the full set of the  numbers describing the system state. Variable $P$ is a quantum number. It is quantizing in a closed edge  the same way as the  non-interacting particles momenta. The distances between the same-spin particles  $a_k,~b_k$ and the quantity $c$ are the classical variables. This can be seen from the Hamiltonian which corresponds to the limit $\hbar\to 0$. The set of classical variables  have arbitrary values;  they have infinite masses and are resting. To some extent, this situation reminds the molecular systems where the electron coordinates are quantum quantities while the ion coordinates are classical. It is known that the molecular system can be considered via the molecular terms: the electronic levels are determined at fixed arbitrary positions of the ions, while the motion of the latter is considered classically where the terms, play the role of the interaction potential. (This description is limited by  crossing of the molecular terms). The relative momentum $P$ is a  global variable.
Thus, the quantities $a_k,~b_k$ and $c$ obey the Boltzmann statistics. This explains how to make average of the observable quantities.

\subsection*{Scattering of interacting electrons at a magnetic impurity}
The exact solution  permits one to include perturbingly other interaction mechanisms that can affect the responses, for example, an interaction with  magnetic impurities or the spin-orbit interaction with phonons.  Here,  as an example, we  consider  backscattering  of the electrons by a magnetic impurity. The backscattering is forbidden without such an interaction  violated the time-reversibility. The e-e interaction essentially modifies the magnetic impurity scattering.
The transition between the states occurs at the terms crossing points (when the total energies of two states are equal at coinciding electron positions).

Consider an impurity whose spin $\bf S$  interacts with the electron spin ${\bm\sigma}_i/2$. The  Hamiltonian of spin-spin interaction is
\begin{equation}\label{s-s}
H_{ss}=U_0\sum_n \delta(x_n)({\bf S}{\bm\sigma}_n)
\end{equation}
For the two-electron system  we find the matrix elements $M=U_0/2e^{i(P+P')a}\delta(a-2b)$  and the transition rate $T=U_0^2/4\delta[E-(P'-P)v-\delta V]$. Here $\delta V=V(a)-\int_0^LV(x)\,dx$ is a correction to the total energy due to interaction. In the general case, the matrix element between the wave functions (\ref{rightleft u}) is equal to $M=U_0/2\sum_{ij}e^{i(P+P')a_{ij}}\delta(a_{ij}-2b)$ and $T=NU_0^2/4\delta[E-(P'-P)v-\delta V]$, where $\delta V=\sum_{ij}V_{ij}-(n-m+1)\int_0^LV(x)\,dx$ includes correction to the interaction energy $V_{ij}$ after transition of the electron from left- to rightmover ensemble or vice-versa.

Now study the many-body problem. Consider the transition between the states $|1\rangle=|P,c,\{a_1,.. a_{n-1}\},\{b_1,...b_{m-1}\}\rangle$ and $|2\rangle=|P',c',\{a_2,...a_{n-1}\},\{a_1,b_1,...b_{m-1}\}\rangle$. The backscattering rate  at fixed quantum numbers is
\begin{eqnarray}\label{scatt}\nonumber
&&2\pi |\langle 1|H_{ss}|2\rangle|^2\times\nonumber\\&&\delta(vP-vP'+\sum_{j=1}^{n-1}U(a_j)-\sum_{j=1}^{m-1}U(b_j))\nonumber=\\&&2\pi|U_0|^2
\delta(vP-vP'+\sum_{j=1}^{n-1}U(a_j)-\sum_{j=1}^{m-1}U(b_j))
\nonumber\\\times&&\sum_{j,k}\delta(a_j+2c)\delta(b_k+2c')
\end{eqnarray}
Averaging with respect to $c,~c'$ gives
\[
\langle\sum_{j,k}\delta(a_j+2c)\delta(b_k+2c')\rangle=(n-1)(m-1)/L^2.
\]
The physical meaning of Eq.(\ref{scatt}) is simple. An electron changing its spin simultaneously changes its interaction energy with all electrons with the same spins to the interaction energy with the opposite-spin-electrons. This difference of potential energies is transmitted to the difference of the kinetic energies establishing the thermal equilibrium between the kinetic and potential energies.

 Classical variables obey the Boltzmann statistics in the thermal equilibrium.  Let us now average the energy delta-function over this distribution:
\begin{eqnarray}\label{1000}
R=\frac{\int\prod da_idb_j \exp(-\beta\tilde{U})\delta(...)}{\int\prod da_idb_j\exp(-\beta\tilde{U})}.
\end{eqnarray}
In the nearest-neighbor approximation  we obtain
\begin{widetext}
\[
  R=\int_{-\infty}^\infty\frac{dt}{2\pi}e^{itv(P-P')}\exp\left\{\frac{1}{L}\int da\left[ n(e^{(it-\beta) U(a)}-1)+m(e^{(-it-\beta) U(a)}-1)\right]\right\}\biggm/\exp\left\{\frac{n+m}{L}\int da\left[e^{-\beta U(a)}-1\right]\right\}\nonumber
\]
\end{widetext}
Eq. (\ref{1000}) gives a symmetric dependence of the transition probability on $P-P'$, which is determined by the function $U(a)$. A careful examination goes beyond the scope of the paper.
\subsection*{Conclusions}
In conclusion, we have found the exact solution of the  interacting many-particle 1D system   with the  linear single-particle spectrum. The Hamiltonian includes an external potential (e.g., impurities) as well.  The Schr\"edinger equation  solution is reduced to the separation of the system to the groups of right- and left-moving carriers with the constant velocity. The interactions between groups and with impurities are reflected in the phase factor in the wave function. The relative coordinates in each group turn out to be classical conserving  variables, while the relative momentum of all carriers is a global quantum number. The collective selfenergy consists of the linear in momentum kinetic part and the potential energy of interaction at a fixed interparticle distance inside the groups. In the framework of the Hamiltonian Eq.~(\ref{100}) the backscattering is absent.
 The formal solution is applicable to the edge states of the 2D topological insulator.
 In a separate paper \cite{hyper} we have found that the edge states  have either an exact linear single-electron spectrum in the most models of the 2D topological insulator  or this spectrum is numerically linear. Hence, the results  of the present paper directly pertain to these edge states.

Unlike the Luttinger liquid, our solution is not linked to the Fermi level of the non-interacting system. It is also valid in a strongly non-equilibrium situation.

The conservation of the distances  between the same-spin electrons makes   relaxation of such  system to the equilibrium impossible, unless some additional term are taken into account.

In accordance with the obtained equations the implementation of the  e-e and electron-impurities interactions has no effect on the velocity matrix elements. That means that the conductivity of the system  is also not changed and stay infinite for the system of interacting electrons.

The exact solutions  permitted one to include perturbingly the interaction with the magnetic impurities that was not included in the Hamiltonian (\ref{100}). It was found that the e-e interaction  essentially affects the backscattering. Note that  other mechanisms, like the spin-orbit interaction with phonons, can be studied in the same way.

One other remark concerns  possibility of the generalization of the Hamiltonian (\ref{100}). It obviously can be generalized to
\begin{eqnarray}\label{1010}
  H=\sum_i (v\sigma_ip_i+U_i(x_i)+\sigma_iU_i^{(1)}(x_i))+\nonumber \\\sum_{i<j}(V_{ij}(x_i-x_j)+\sigma_i\sigma_jV_{ij}^{(1)}(x_i-x_j)),
\end{eqnarray}
with similar consequences.

\subsection*{Acknowledgements}
This research was supported  by RFBR grant  No  17-02-00837. The authors thank A.V. Chaplik for stimulating discussions.

\end{document}